\documentclass[conference,10pt]{IEEEtran}
\IEEEoverridecommandlockouts
\usepackage{cite}
\usepackage{amsmath,amssymb,amsfonts}
\usepackage{algorithmic}
\usepackage{graphicx}
\usepackage{textcomp}
\usepackage{xcolor}
\def\BibTeX{{\rm B\kern-.05em{\sc i\kern-.025em b}\kern-.08em
    T\kern-.1667em\lower.7ex\hbox{E}\kern-.125emX}}

\usepackage[acronym]{glossaries}
\newacronym{RIS}{RIS}{Reconfigurable Intelligent Surfaces}

\begin{document}

\title{Identification of RIS-Assisted Paths for Wireless Integrated Sensing and Communication\\
\thanks{The financial support by the Austrian Federal Ministry for Digital and Economic Affairs, the National Foundation for Research, Technology and Development, and the Christian Doppler Research Association is gratefully acknowledged.}
}

\author{\IEEEauthorblockN{
Zeyu Huang\IEEEauthorrefmark{1}\IEEEauthorrefmark{2}, \textit{Student Member, IEEE},   % 1st author, 1st affiliations
Stefan Schwarz\IEEEauthorrefmark{1}\IEEEauthorrefmark{2}, \textit{Senior Member, IEEE},   % 2nd author, 2nd affiliations
Bashar Tahir\IEEEauthorrefmark{1}\IEEEauthorrefmark{2}, \\  % 3rd author, 3rd affiliations
Markus Rupp\IEEEauthorrefmark{1}, \textit{Fellow, IEEE},    % 
}                                     % ...
%\\
\IEEEauthorblockA{\IEEEauthorrefmark{1}% 1st affiliations
Institute of Telecommunications, TU Wien, Vienna, Austria,} 
\IEEEauthorblockA{\IEEEauthorrefmark{2}% 2nd affiliations
 Christian Doppler Laboratory for Dependable Wireless Connectivity for a Society in Motion.}
E-mails: \{zeyu.huang, stefan.schwarz, bashar.tahir, markus.rupp\}@tuwien.ac.at}


\maketitle

\begin{abstract}
Distinguishing between reconfigurable intelligent surface (RIS) assisted paths and non-line-of-sight (NLOS) paths is a fundamental problem for RIS-assisted integrated sensing and communication. In this work, we propose a pattern alternation scheme for the RIS response that uses part of the RIS as a dynamic part to modulate the estimated channel power, which can considerably help the user equipments (UEs) to identify the RIS-assisted paths. Under such a dynamic setup, we formulate the detection framework for a single UE, where we develop a statistical model of the  estimated channel power, allowing us to analytically evaluate the performance of the system. We investigate our method under two critical factors: the number of RIS elements allocated for the dynamic part and the allocation of RIS
elements among different users. Simulation results verify the accuracy of our analysis.

\end{abstract}


\section{Introduction}

Next-generation communication systems are not
only expected to provide powerful communication capabilities, but also to have an environment-aware ability to support novel applications, such as Internet of Things (IoT) services\cite{b1,b11}, autonomy\cite{b2}, and crowdsensing\cite{b3}. To realize this vision, reconfigurable intelligent surfaces (RISs) might be essential for integrating sensing and communication due to their ability to shape the propagation environment\cite{b4,b5,bashar,le}. 

Efforts have been taken in the literature to study RIS-aided positioning.
In \cite{b6}, the authors derived the Cram\'er-Rao lower bound (CRLB) 
of a general RIS-aided localization scenario, and also proposed a spherical wavefront closed-form RIS phase design. 
In \cite{b7}, the authors analyzed localization performance of RISs for both geometric near-field (NF) and far-field (FF) regimes.
In\cite{b8}, a two-stage RIS-aided localization algorithm has been proposed, where the time-of-arrival (ToA) and the direction-of-arrival (DoA) of reference signals have been used to estimate the position. 
The impact of RIS characteristics on the positioning performance, such as the phase shift configuration and the number of RIS elements, have been studied in
\cite{b9,b10}.

 %The estimation methods to obtain these characteristics and localization algorithm based on them have been studying for many years. Such as space-alternating generalized expectation-maximization algorithm (SAGE)\cite{b5} and compressing sensing\cite{b6} have been proposed to estimated the parameters. Time of Arrival (TOA) , Time Difference of Arrival (TDOA), or the Angle of Arrival (AOA)\cite{b3,b4} have been used to estimate the position information. However, these localization methods in the aforementioned works are usually ask for the information of paths. These method have also been extended to RIS-aided localization scenario, such as

\begin{figure}[t]
\centerline{\includegraphics[width=0.35\textwidth]{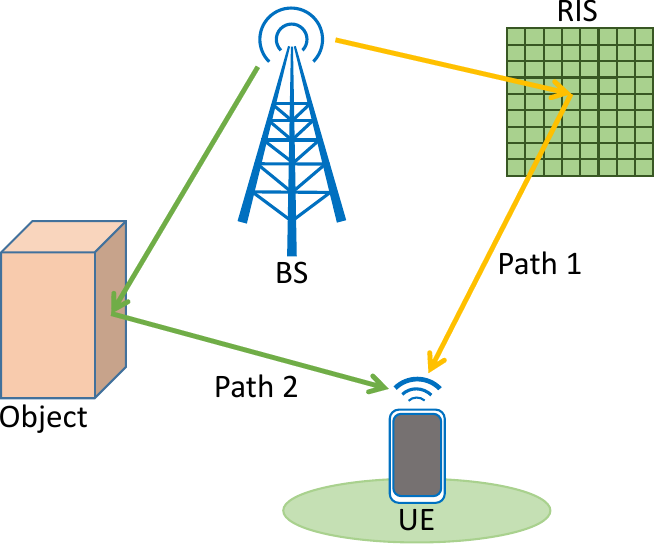}}
\caption{A UE receiving signals from two paths over the RIS and the scattering object, respectively.}
\label{figintro}
\end{figure}

\acrshort{RIS} is usually employed as a reference point in the works mentioned above. The characteristics of RIS-assisted multiple propagation paths play a crucial role, such as delay, Doppler shift, angle information and power. However, in a practical environment, in addition to the propagation path from \acrshort{RIS}, there are also non-line-of-sight (NLOS) paths caused by other objects in the environment..  
As shown in Fig.\,\ref{figintro}, the user equipment (UE) will receive reflections not only
from the RIS but also from other scattering objects in
the environment. 
This path from another scattering
object endangers successful positioning based on the
RIS-assisted path.

In this study, we propose a RIS configuration pattern to help identify the RIS-assisted path by modulating the estimated channel power. 
After we obtain the characteristics of the RIS-assisted path, such as ToA or DoA, combining these characteristics with the prior information on the
RISs positions, we can estimate the location of the UEs. In our pattern configuration, part of the RIS elements are alternatingly configured for a coherent and a random combining of the UEs' signals. We refer to these elements as the dynamic part of the RIS. With the help of the dynamic part, UEs can distinguish the RIS-assisted path based on the estimated channel power. 
The fundamental problem in our method is the detection of the estimated channel power of the two alternating coherent and random patterns of the dynamic part. 

The main contributions of this paper are: 
\begin{itemize}
    \item We propose a scheme which takes advantage of the ability of RIS to shape the propagation channel, to identify the RIS-assisted path from NLOS paths.
    \item We formulate a detection framework for a single UE and develop a statistical model of the estimated channel power.
    \item We investigate the impact of two critical factors under our detection framework: the number of RIS elements allocated for the dynamic part and the allocation of RIS elements among different users.
\end{itemize}

The paper is organized as follows. We first introduce the system model and define the detection problem and performance metrics in Section \ref{systemmodel}. The derivation of the statistical model of the estimated channel power is elaborated in Section \ref{statistial}. We investigate the performance of our scheme in Section \ref{performance}. Finally, we give concluding remarks in Section V.

\section{RIS-Assisted Path Identification}\label{systemmodel}
\subsection{System Model}
\begin{figure}[t]
\centerline{\includegraphics[width=0.4\textwidth]{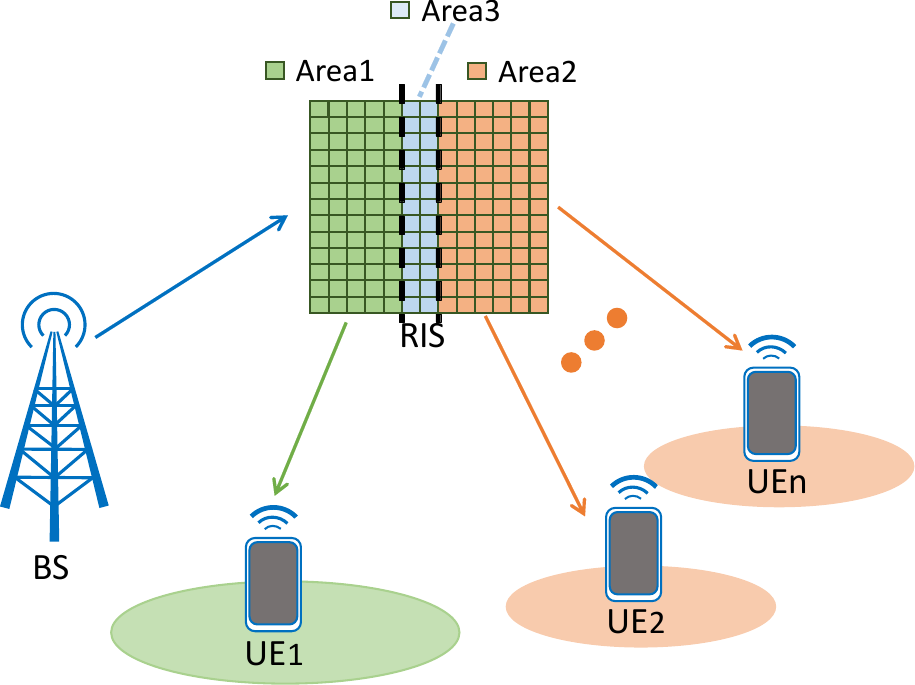}}
\caption{Scenario of our system model.}
\label{fig0}
\end{figure}

We consider a downlink consisting of a single-antenna base station (BS) and $n$ single-antenna UEs. 
We assume the direct LOS paths between the BS and the UEs are blocked, and therefore the communication takes place primarily through the RIS. 
The BS transmits a training signal for sensing. 
Without loss of generality, we will focus here on UE\,1.
In our setup, the RIS elements are divided into three regions: area\,1, area\,2 and area\,3, as Fig.\, \ref{fig0} shows. 
For UE\,1, the received  signal consists of three parts. The first part is from area\,1, where the phase shifts of the RIS elements are configured to combine the training signal to UE\,1 coherently. The second part is the signal from area\,2, which is configured to combine the training signal to the other UEs coherently. Area\,3 is the dynamic part, which is alternatingly configured for coherent combination to one of the UEs, one at a time. 

The received signal $y$ for UE\,1 is expressed by 
\begin{equation}
\begin{split}
     y = \mathbf{h}_{\mathrm{r}}^{\text{T}}\mathbf{\Omega}\mathbf{h}_{\mathrm{t}}x+m.
\end{split}         
\end{equation}
Where $\mathbf{h}_{\mathrm{r}}$ and $\mathbf{h}_{\mathrm{t}}$ are the channels from \acrshort{RIS} to receiver, and from transmitter to \acrshort{RIS}, respectively,  $x$ is the transmitted training signal, and $m$ is the complex Gaussian noise. The phase shift matrix of the \acrshort{RIS} $\mathbf{\Omega}$ is given by 
\begin{equation}
         \mathbf{\Omega} = \text{diag}\left(\mathbf{h}_{\mathrm{RIS}} \right),
\end{equation}
\begin{equation}         \mathbf{h}_{\mathrm{RIS}} = \left( e^{j\beta_{1}},e^{j\beta_{2}}, \dots, e^{j\beta_{\mathrm{Q}}}  \right)^{\text{T}},
\end{equation}
where $Q$ is the total number for \acrshort{RIS} elements and $\beta_{q}$ is the phase shift applied at the $q$-th \acrshort{RIS} element.

The effective channel $\mathbf{h}_{\mathrm{r}}^{\text{T}}\mathbf{\Omega}\mathbf{h}_{\mathrm{t}} $ has three parts
\begin{equation}
\begin{split}
     \mathbf{h}_{\mathrm{r}}^{\text{T}}\mathbf{\Omega}\mathbf{h}_{\mathrm{t}} 
       &= \sum_{q \in \mathcal{A}_1}h_{\mathrm{r},q} h_{\mathrm{t},q}e^{j\beta_{q}} + \sum_{q \in \mathcal{A}_2}h_{\mathrm{r},q} h_{\mathrm{t},q}e^{j\beta_{q}} \\&+\sum_{q \in \mathcal{A}_3}h_{\mathrm{r},q} h_{\mathrm{t},q}e^{j\beta_{q}},
\end{split}  
\label{receivedsingal}
\end{equation}
where $\mathcal{A}_1$, $\mathcal{A}_2$, and $\mathcal{A}_3$ are the set of RIS elements corresponding to area\,1, area\,2, and area\,3 of the RIS, respectively. The number of elements for the corresponding parts are $N = |\mathcal{A}_1|$, $M = |\mathcal{A}_2|$, and $K = |\mathcal{A}_3|$, respectively.  
%As mentioned before,  $\mathcal{A}_3$ corresponds to the dynamic part of the RIS, which either coherently combines the signals when it is configured for the \ac{UE}1 or randomly when it is configured for other UEs.
The coefficients $h_{\mathrm{r},q}$ and $ h_{\mathrm{t},q}$ represent
the channel from the $q$-th \acrshort{RIS} element to UE\,1 and the channel from BS to the $q$-th \acrshort{RIS} element, respectively. We consider the free-space pathloss channel model here. For example, the channel $h_{\mathrm{r},q}$ is modeled by
\begin{equation}
         h_{\mathrm{r},q} = \frac{\lambda}{4\pi r_{\mathrm{r},q}}\exp\left({-j\frac{2\pi}{\lambda}r_{\mathrm{r},q}}\right),
\end{equation}
where $r_{\mathrm{r},q}$ is the distance between the $q$-th \acrshort{RIS} element and the UE and $\lambda$ is the wavelength.

\subsection{Problem Formulation}
\label{parmaters}
The dynamic part is alternatingly configured among the users.
Hence, for UE\,1, two patterns exist:
\begin{itemize}
    \item pattern\,1, the dynamic part is configured to combine  the signal to UE\,1 coherently;
    \item pattern\,2, the dynamic part is configured for another UE.
\end{itemize}
For UE\,1, the power of the estimated channel is higher in pattern\,1 compared to that of pattern\,2. We use these two patterns to encode the dynamic part.
By detecting the power variation of the RIS-assisted path, 
we can distinguish between the RIS-assisted path and other NLOS paths, since only the signal coming from the RIS-assisted path would have such an alternating power level in time.

As we are interested in detecting the presence of these two alternating patterns, we define a hypothesis problem as
\begin{equation*}
\mathrm{H_{1}}:\hat{h} = h_{1},
\end{equation*}
\begin{equation*}
\mathrm{H_{2}}:\hat{h} = h_{2},
\end{equation*}
where $h_{1}$ is the effective channel resulting from pattern\,1, and $h_{2}$ is the effective channel resulting from pattern\,2, and $\hat{h}$ is the estimated channel at UE\,1.

In order to evaluate the performance of the correct pattern detection, 
we define the detection error probability as 
\begin{equation}
    \begin{split}
          \mathrm{P}_{\mathrm{error}} &= \mathrm{P} \left \{ \mathrm{H_{1}} \right \}  \mathrm{P} \left \{  \left |\hat{h}\right |^2 < \gamma \bigg|\mathrm{H_{1}} \right \}\\&+\mathrm{P} \left \{ \mathrm{H_{2}} \right \}  \mathrm{P} \left \{ \left |\hat{h}\right |^2 > \gamma \bigg|\mathrm{H_{2}} \right \}.
    \end{split}
    \label{Perror}
\end{equation}
where $\gamma$ is a threshold, and $\mathrm{P} \left \{  \left |\hat{h}\right |^2 < \gamma \bigg|\mathrm{H}_{i} \right \}$ is the cumulative distribution function (CDF) of $ \left|\hat{h}\right |^2$ under hypothesis $\mathrm{H}_{i}$, $i\in \left \{ 1,2  \right \}$.
In this study, we have $ \mathrm{P} \left \{ \mathrm{H_{1}} \right \} =  \mathrm{P} \left \{ \mathrm{H_{2}} \right \} =0.5$, and $\gamma$ is set to 
\begin{equation}
\gamma= \arg\min \limits_{\mu_{2}<x<\mu_{1}}\mathrm{P}_{\mathrm{error}},
\end{equation}
where $\mu_{i}=\mathrm{E}\left \{ \left |\hat{h}\right |^2\bigg|\mathrm{H_{i}} \right \}, i\in \left \{ 1,2  \right \}$.
That is, we set the threshold to minimize the detection error probability.

The more elements are allocated for the dynamic part, the more RIS elements will participate in the pattern\,2 phase, which in turn reduces the total received power for UE\,1 during that phase. Hence, the number of RIS elements that are used in the dynamic part is a crucial parameter. In order to capture the impact of this, we define the random part ratio $R$ as the percentage of elements that are used in the dynamic part, and which would result in random combining during the phase of pattern 2 for UE\,1. This is given by
\begin{equation}
    R = \frac{K}{N+M+K}.
\end{equation}
We define the relative power difference as 
\begin{equation}
        G _{\mathrm{d}}=10\log_{10}
        \frac{\mathrm{E}\left \{ \left |\hat{h}\right |^2\bigg|\mathrm{H_{1}} \right \} }{\mathrm{E}\left \{ \left |\hat{h}\right |^2\bigg|\mathrm{H_{2}} \right \} }.
        \label{Gloss}
\end{equation}
This allows us to tell the impact of dynamic switching patterns on communication link quality for UE\,1. 
%We calculate the deflection of our hypotheses problem as
%\begin{equation}
%    d^{2} =  \frac{\left(E \left \{ \left |\hat{h}_{1}   \right |^2 \bigg|\mathrm{H_{1}} \right \} - E \left \{ \left |\hat{h}_{2}   \right |^2\bigg|\mathrm{H_{2}} \right \}\ \right)^2}{var\left \{ \left |\hat{h}\right |^2\bigg|\mathrm{H_{1}} \right \}  },
%    \label{deflection}
%\end{equation}
%which is a metric to evaluate the distance between the distributions from different hypotheses. We will use deflection later to explain the impact on detection error probability and power difference  when we use different RIS models.  

\section{Statistical Characteristics of Received Power}
\label{statistial}

\subsection{Received Signal from Different RIS Areas}
First, we study the received signal of UE\,1 from area\,1. For
an ideal RIS model, it is assumed that the phase shifts can
span the entire range of $2\pi$, and no amplitude attenuation
occurs. In this case, perfect coherent combing is possible,
where the signals from all propagation paths are aligned
in phase:
\begin{equation}
         e^{-j\beta_{q}} = \frac{h_{\mathrm{r},q} h_{\mathrm{t},q}}{\left |h_{\mathrm{r},q}  \right | \left | h_{\mathrm{t},q}  \right |}.
\label{equation4}
\end{equation}
Hence, we can express the signal from area\,1 in equation \eqref{receivedsingal} as
\begin{equation}
    \sum_{q \in \mathcal{A}_1}h_{\mathrm{r},q} h_{\mathrm{t},q}e^{j\beta_{q}}= \sum_{q\in \mathcal{A}_1}\left |h_{\mathrm{r},q}  \right | \left | h_{\mathrm{t},q}  \right |.
\end{equation}

We assume that area\,2, which is configured for the other users, keeps switching in configuration for the other users over time. 
Hence, for UE\,1, the phase shifts of the RIS elements
in area\,2 can be considered as random variables, with $~\beta_{q}\sim U\left (-\pi,\pi  \right )$. In this study, we assume that all the channel coefficients have similar amplitude. 
%which means the distance from RIS to UE\,1 is much larger than the size of RIS. 
Hence, with a sufficient number $M$, based on the central limit theorem, $\sum_{q\in \mathcal{A}_2}h_{\mathrm{r},q} h_{\mathrm{t},q}e^{j\beta_{q}}$ can be considered as a complex Gaussian random variable,
\begin{equation}
    \sum_{q\in \mathcal{A}_2}h_{\mathrm{r},q} h_{\mathrm{t},q}e^{j\beta_{q}} \sim \mathcal{CN}\left (0,\sum_{q \in \mathcal{A}_2}\left |h_{\mathrm{r},q}   h_{\mathrm{t},q}  \right |^2\right).
\end{equation}
The proof of this can be found in Appendix \ref{appA}.

For the signal from area\,3, if pattern\,1 is active, we have 
\begin{equation}
    \sum_{q \in \mathcal{A}_3}h_{\mathrm{r},q} h_{\mathrm{t},q}e^{j\beta_{q}}= \sum_{q\in \mathcal{A}_3}\left |h_{\mathrm{r},q}  \right | \left | h_{\mathrm{t},q}  \right |.
\end{equation}
If pattern\,2 is active in area\,3, then we have a similar setup as in area\,2, where we can use a complex Gaussian to characterize the received signal. This is given by
\begin{equation}
    \sum_{q\in \mathcal{A}_3}h_{\mathrm{r},q} h_{\mathrm{t},q}e^{j\beta_{q}} \sim \mathcal{CN}\left (0,\sum_{q \in \mathcal{A}_3}\left |h_{\mathrm{r},q}   h_{\mathrm{t},q}  \right |^2\right).
\end{equation}

\subsection{Estimated Channel}
We can estimate the channel by $\hat{h}=\frac{y}{x}$.
Hence, the estimated  channels of pattern\,1 and pattern\,2 for UE\,1 are expressed by
\begin{equation}
\begin{split}
        \hat{h}_{1} &= \sum_{q\in \mathcal{A}_1}\left |h_{\mathrm{r},q}  \right | \left | h_{\mathrm{t},q}  \right |+\sum_{q\in \mathcal{A}_2}h_{\mathrm{r},q} h_{\mathrm{t},q}e^{j\beta_{q}}
        \\
        &+\sum_{q\in \mathcal{A}_3}\left |h_{\mathrm{r},q}  \right | \left | h_{\mathrm{t},q}  \right |+n,
\end{split}
\end{equation}
and
\begin{equation}
\begin{split}
       \hat{h}_{2}
       &= \sum_{q\in \mathcal{A}_1}\left |h_{\mathrm{r},q}  \right | \left | h_{\mathrm{t},q}  \right|+\sum_{q\in \mathcal{A}_2}h_{\mathrm{r},q} h_{\mathrm{t},q}e^{j\beta_{q}}\\
       &+\sum_{q\in \mathcal{A}_3}h_{\mathrm{r},q} h_{\mathrm{t},q}e^{j\beta_{q}}  +n,
\end{split}   
\end{equation}
respectively,
where $n=\frac{m}{x}$ is a complex Gaussian random variable, $n \sim \mathcal{CN}\left (0,\sigma^{2}_{\mathrm{N}}\right)$.
With the assumption that the random phase configurations for each RIS element and the noise are independent, we can conclude that both $\hat{h}_{1}$ and $\hat{h}_{2}$ are complex Gaussian random
variables.

\subsection{Power of the Estimated Channel}
%$\hat{h}_{1}$ and $\hat{h}_{2}$ have 
%We denote here the estimated channel as $\hat{h}$, which $\hat{h} \in \{\hat{h}_{1},\hat{h}_{2}\}$.
The power of estimated channel $\hat{h}$ is expressed by 
\begin{equation}
    \left |\hat{h}_{i}\right |^2 =  \left |\Re\{\hat{h}_{i}\}\right |^2+\left |\Im\{\hat{h}_{i}\}\right |^2, \ i\in \left \{ 1,2  \right \},
\end{equation}
where $\Re\{\hat{h}_{i}\}$ and $\Im\{\hat{h}_{i}\}$ are Gaussian random variables based on the central limit theorem, and they have the same variances. The proofs are in Appendix \ref{appB}. 

In Appendix \ref{appC}, we show that
\begin{equation}
    \mathbf{E}\left\{\Re\{\hat{h}_{i}\} \Im\{\hat{h}_{i}\}\right \} = 0, \ i\in \left \{ 1,2  \right \}.
\end{equation}
Due to the $\hat{h}_{i}$ being complex Gaussian random variables, and $\Re\{\hat{h}_{i}\}$ and $\Im\{\hat{h}_{i}\}$ being independent,
we can use a two degree of freedom non-central chi-squared distribution to characterize the sum of squares of two independent Gaussian random variables with non-zero means and unit variances. 
Hence, we have non-central chi-squared distributions for pattern\,1 and pattern\,2 as
\begin{equation}
    \frac{1}{\sigma^{2}_{\mathrm{1}}}\left |\hat{h}_{1}   \right |^2 \hskip 1pt \sim  \hskip 1pt \chi^{2}  \left (2,\frac{1}{\sigma^{2}_{\mathrm{1}}}\left |\sum_{q\in \mathcal{A}_1}\left |h_{\mathrm{r},q}  \right | \left | h_{\mathrm{t},q} \right | +  \sum_{q\in \mathcal{A}_3}\left |h_{\mathrm{r},q}  \right | \left | h_{\mathrm{t},q} \right | \right |^2  \right),
    \label{nccs1}
\end{equation}
and
\begin{equation}
    \frac{1}{\sigma^{2}_{\mathrm{2}}}\left |\hat{h}_{2}   \right |^2  \sim \chi^{2} \left (2,\frac{1}{\sigma^{2}_{\mathrm{2}}} \left |\sum_{q\in \mathcal{A}_1}\left |h_{\mathrm{r},q}  \right | \left | h_{\mathrm{t},q}  \right | \right |^2 \right),
    \label{nccs2}
\end{equation}
where  
\begin{equation}
    \sigma^{2}_{\mathrm{1}} =\frac{1}{2} \sigma^{2}_{\mathrm{N}}++ \frac{1}{2} \sum_{q\in \mathcal{A}_2}\left |h_{\mathrm{r},q}   h_{\mathrm{t},q}\right |^2,
    \label{nccs3}
\end{equation}
\begin{equation}
    \sigma^{2}_{\mathrm{2}} = \frac{1}{2} \sigma^{2}_{\mathrm{N}}+ \frac{1}{2} \sum_{q\in \mathcal{A}_2}\left |h_{\mathrm{r},q}   h_{\mathrm{t},q}\right |^2+ \frac{1}{2} \sum_{q\in \mathcal{A}_3}\left |h_{\mathrm{r},q}   h_{\mathrm{t},q}\right |^2.
    \label{nccs4}
\end{equation}

\section{Performance Analysis}
\label{performance}
\subsection{Simulation Scenario}
We compare our analysis with simulations. We have a 2-
dimensional scenario containing a single-antenna BS, a single-antenna UE, and a uniform linear array (ULA) RIS.
In the Cartesian coordinate system, the coordinates of the BS are (0\,m, 0\,m) and the center of RIS array are (25\,m, 25\,m), the free space pathloss between BS and RIS is about 77\,dB. The power of noise is about -132\,dBm.
The common parameters are summarized in Table \ref{table:table1}. We give other parameters in the relevant subsection later.

\begin{table}
	\centering
	\caption{Simulation parameters.\label{table:table1}}
	\begin{small}
		\begin{tabular}{cc}
        \hline
			Bandwidth &$15$\,KHz\\[0.5ex]
			\hline
			Central frequency & 5\,GHz \\[0.5ex]
			\hline
			Noise power density  &-174\,dBm/Hz\\[0.5ex]
			\hline
			BS transmit power  &30\,dBm\\[0.5ex]
			\hline
			RIS elements number & 1000 \\
            \hline
			RIS elements spacing & half wavelengths \\
            \hline
		\end{tabular}
	\end{small}
\end{table}

\subsection{Statistical Characteristics}
% Statistical characteristics of the estimated channel power are fundamental to our detection problem, because we need them to calculate the detection error probability and relative power difference. 

\begin{figure}[t]
\centerline{\includegraphics[width=0.5\textwidth]{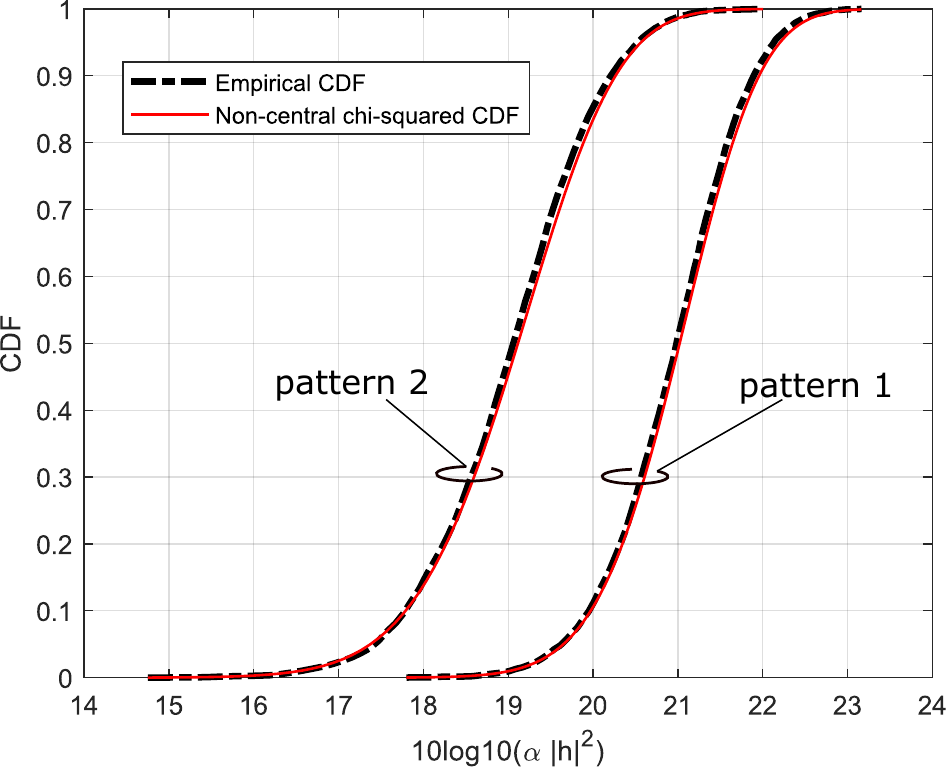}}
\caption{Empirical and analytical CDFs. In pattern\,1, $\alpha = \frac{1}{\sigma^{2}_{\mathrm{1}}}$, $h = \hat{h}_{1}$. In pattern\,2, $\alpha = \frac{1}{\sigma^{2}_{\mathrm{2}}}$, $h = \hat{h}_{2}$. $N=500, M=400$ and $K=100$.}
\label{fig4}
\end{figure}

Fig.\,\ref{fig4} shows an example of the empirical and analytical CDFs of $\frac{1}{\sigma^{2}_{\mathrm{1}}}\left |\hat{h}_{1}   \right |^2$ and $\frac{1}{\sigma^{2}_{\mathrm{2}}}\left |\hat{h}_{2}   \right |^2$, which are from pattern\,1 and pattern\,2, respectively. The empirical CDFs are represented by the black dashed line, and the analysis CDFs are the solid red line.
The RIS elements in area\,1, area\,2 and area\,3 are $N=500$, $M=400$, and $K=100$. The Cartesian coordinates of the UE\,1 are (7000\,m, 0\,m).
% The analytical and simulated non-centrality parameters of the non-central chi-squared distribution for pattern\,1 are 125.49 and 123.36, respectively. For pattern\,2, the values are 80.55 and 81.67. 
We can see that our analysis matches nicely with simulation. 

\subsection{Impact of $R$}

The impact of our dynamic switching pattern on communication is a crucial problem, primarily when the UE communication link is mainly supported by the RIS. This means we need to make a trade-off between path identification and communication. The random part ratio is the crucial parameter that must be studied to achieve this balance. The analytical detection error probability and relative power loss are obtained by evaluating \eqref{Perror} and \eqref{Gloss} based on the expressions of non-central chi-squared distribution in \eqref{nccs1} to \eqref{nccs4}.

\begin{figure}[t]
\centerline{\includegraphics[width=1\linewidth]{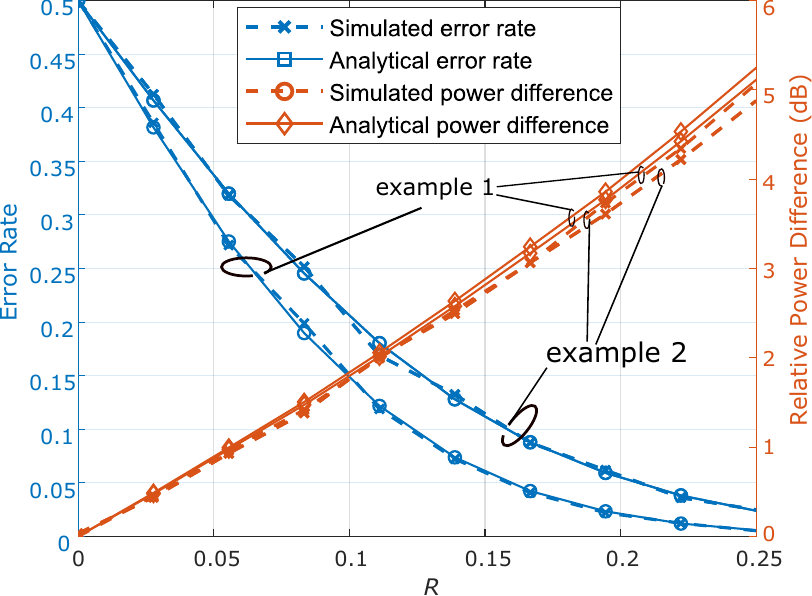}}
\caption{Detection error probability and relative power loss w.r.t. $R$. $N + K = 600, M=400$. With the increases of $R$, $K$ increases and $N$ decreases.}
\label{fig5}
\end{figure}

Fig.\,\ref{fig5} shows the detection error probability and relative power difference w.r.t. random part ratio.
The x-axis is the random part ratio $R$ as we defined before. The left y-axis is the detection error probability, and the right y-axis is the relative power difference. We have two examples in Fig.\,\ref{fig5}. The UE coordinates are (7000\,m, 0\,m) in example\,1, and (9000\,m, 0\,m) in example\,2, and the pathloss between RIS and UE are  about 123\,dB and 125\,dB. respsctivly. The number of RIS elements in area\,1 and area\,3 is $N+K=600$, and in area\,2 is $M=400$. With the increase of $R$, $K$ increases and $N$ decreases.

We can see that with the increase in the $R$, the detection error probability for our pattern sensing scheme gradually decreases, and its decreasing rate gradually becomes slower. Meanwhile, the relative power loss increases.
Example\,1 corresponds to a lower detection error probability for a given $R$. The relative power slightly increases. This is because the CDFs of the estimated channel power in example\,1 are more separate than in example\,2. Which will cause a larger relative power difference. For instance, we can see that in order to decrease the detection error probability to 0.1, the $R$ need to be about 0.125 in example\,1. In example\,2, to achieve the same detection error probability, we need $R$ equal to about $0.16$. 

We can also see that our analysis results are an excellent match with the simulated results.  The maximum difference in detection error probability between analysis and simulation is about 0.01 in example\,1 when $R$ is about 0.08. The maximum difference of relative power difference between analysis and simulation is about 0.23\,dB in example\,2 when $R$ is about 0.25.

\subsection{Impact of the Number of Elements in Area\,2}

For a given random part ratio $R$, the more RIS elements are configured to combine the
signal to UE 1, the fewer RIS elements are available for other UEs. In this subsection, we study the impact of the division of elements between UEs for a given configuration of the dynamic part.

\begin{figure}[t]
\centerline{\includegraphics[width=1\linewidth]{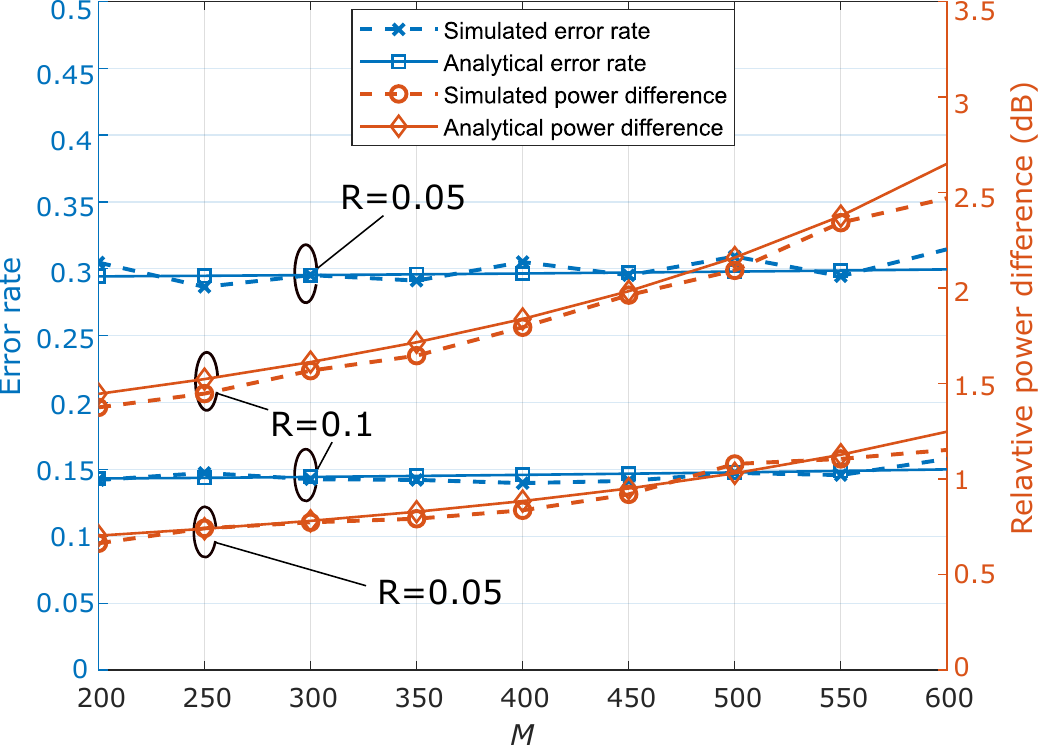}}
\caption{Detection error probability and relative power loss w.r.t. $M$.  When $R=0.1$, we have $N+M=900, K=100$. When $R=0.05$, $N+M=950, K=50$.}
\label{fig9}
\end{figure}

Fig.\,\ref{fig9} shows two examples that have different RIS elements in area\,3.  When $R=0.1$, we have $N+M=900, K=100$. When $R=0.05$, $N+M=950, K=50$. The x-axis is the number of RIS elements in area\,2. With the increase of $M$, $N$ decreases.  The left y-axis is the detection error probability, and the right side is the relative power difference. The UE coordinates are (7000 m, 0 m).

From Fig.\,\ref{fig9}, we can see that, in general, the detection error probability does not change substantially with the increase of $M$, but does with $R$. For a higher $R$, we have a lower detection error probability. The detection error probability difference between the cases $R=0.1$ and $R=0.05$ is about 0.15. 
With the increase of $M$, the relative power difference increases for both the cases of  $R=0.1$ and $R=0.05$, about 1.2\,dB and 0.5\,dB, respectively. We can see that with a larger $R$, we can have a lower detection error probability. However, we will have a larger relative power difference.

Our analysis results are still an excellent match with the simulated results.
The maximum difference in detection error probability between analysis and simulation is about 0.012 when $M=500$ and $R=0.05$. The maximum difference of relative power difference between analysis and simulation is about 0.18\,dB when $M=600$ and $R=0.1$.

\section{Conclusions}
\label{conclusion}
In this paper, we considered the problem of RIS path identification, where a dynamic RIS configuration scheme is proposed. 
In our scheme, we used part of RIS elements as the dynamic part, which is configured in an alternating fashion to coherently combine the signals to the UEs, one at a time. Hence, the UEs can distinguish the RIS-assisted path based on a designed pattern of the estimated channel power. 
We formed a hypothesis detection framework for a UE to detect the active pattern in the dynamic part of the RIS. We carried out a statistical analysis of the problem, where we showed how non-central chi-squared distribution can be used to model the estimated channel power at the UE.
We studied the impact of the number of RIS elements allocated for the dynamic part. We concluded from analysis and simulation that more RIS elements in the dynamic part could help the UE detect different patterns in the RIS dynamic part. However, as the sensing detection error probability decreases, the total received power of the UE decreases as well, which in turn affects the communication performance.
We found that the detection error probability for a single UE is mainly subjected to the size of the dynamic part.  However, a single UE will experience more power loss in our scheme with fewer RIS elements which we configure to combine the signal to it coherently.

\vspace{12pt}

\appendix

\subsection{Statistics of the dynamic part}
\label{appA}
The dynamic part $ h_{d}$ is expressed by
\begin{equation}
\begin{split}
        h_{\mathrm{d}} = \sum_{q=1}^{\mathrm{K}}h_{\mathrm{r},q} h_{\mathrm{t},q} e^{j\beta_{q}}, 
\end{split}
\end{equation}
where  $\beta_{q} \sim U\left (-\pi,\pi \right )$.
Let us make 
\begin{equation}
h_{\mathrm{r},q} h_{\mathrm{t},q}e^{j\beta_{q}} = \left( a_{q}+j b_{q}\right)e^{j\beta_{q}}.
\end{equation}
Then we have 
% \begin{equation}
% \begin{split}
%         h_{\mathrm{d}} &= \sum_{q=1}^{\mathrm{K}}\left(a_{q} \cos{\beta_{q}}- b_{q}\sin{\beta_{q}}\right)\\
%         &+j \sum_{p=1}^{\mathrm{K}}\left(a_{p} \sin{\beta_{p}}+ b_{p}\cos{\beta_{p}}\right).
% \end{split}
% \end{equation}
\begin{equation}
\begin{split}
\mathbf{E}\left\{ \Re\{h_{\mathrm{d}}\} \right\} 
&= \mathbf{E}\left\{ \sum_{q=1}^{\mathrm{K}}\left(a_{q} \cos{\beta_{q}}- b_{q}\sin{\beta_{q}}\right) \right\}\\
% &= \sum_{q=1}^{\mathrm{K}} \mathbf{E}\left\{\left(a_{q} \cos{\beta_{q}}- b_{q}\sin{\beta_{q}}\right) \right\}\\
&= \sum_{q=1}^{\mathrm{K}} \frac{1}{2\pi}\int_{-\pi}^{\pi} \left(a_{q} \cos{\beta}- b_{q}\sin{\beta}\right) \mathrm{d}\beta=0.
\end{split}
\end{equation}
\begin{equation}
\begin{split}
\mathbf{E}\left\{ \Im\{h_{\mathrm{d}}\} \right\} 
= \sum_{p=1}^{\mathrm{K}} \frac{1}{2\pi}\int_{-\pi}^{\pi} \left(a_{q} \sin{\beta}+ b_{q}\cos{\beta}\right) \mathrm{d}\beta=0.
\end{split}
\end{equation}
\begin{equation*}
\begin{split}
&\mathbf{E}\left\{ \Re\{h_{\mathrm{d}}\} \Im\{h_{\mathrm{d}}\} \right\} 
\\&= \mathbf{E}\left\{ \sum_{q=1}^{\mathrm{K}}\left(a_{q} \cos{\beta_{q}}- b_{q}\sin{\beta_{q}}\right)   \sum_{p=1}^{\mathrm{K}} \left(a_{p} \sin{\beta}+ b_{p}\cos{\beta}\right)\right\} \\
&= \mathbf{E}\left\{ \sum_{q=1}^{\mathrm{K}}\sum_{p=1}^{\mathrm{K}} \left(a_{q} \cos{\beta_{q}}- b_{q}\sin{\beta_{q}}\right) \left(a_{p} \sin{\beta}+ b_{p}\cos{\beta}\right)\right\} \\
\end{split}
\end{equation*}
When $q \neq p$, 
\begin{equation*}
\begin{split}
    &\mathbf{E}\left\{ \left(a_{q} \cos{\beta_{q}}- b_{q}\sin{\beta_{q}}\right) \left(a_{p} \sin{\beta}+ b_{p}\cos{\beta}\right)\right\} 
    %\\
    % &= \mathbf{E}\left\{ \left(a_{q} \cos{\beta_{q}}- b_{q}\sin{\beta_{q}}\right) \right\}\mathbf{E}\left\{\left(a_{p} \sin{\beta}+ b_{p}\cos{\beta}\right)\right\} 
    = 0
\end{split}
\end{equation*}
When $q=p$,
\begin{equation*}
\begin{split}
    &\mathbf{E}\left\{ \left(a_{q} \cos{\beta_{q}}- b_{q}\sin{\beta_{q}}\right) \left(a_{p} \sin{\beta}+ b_{p}\cos{\beta}\right)\right\} \\
    &= \frac{1}{2\pi}\int_{-\pi}^{\pi} \left(a_{q} b_{q}\cos^{2}{\beta}- a_{q}b_{q}\sin^{2}{\beta}\right) \mathrm{d}\beta = 0
\end{split}
\end{equation*}
Hence, we have
\begin{equation}
\begin{split}
\mathbf{E}\left\{ \Re\{h_{\mathrm{d}}\} \Im\{h_{\mathrm{d}}\} \right\} 
= 0.
\end{split}
\end{equation}

We need the $\mathbf{E}\left\{ \left(\Re\left\{h_{\mathrm{d}}\right\} \right)^{2}\right\}$ to calculate the variance. 
\begin{equation*}
\begin{split}
\mathbf{E}\left\{ \left(\Re\left\{h_{\mathrm{d}}\right\} \right)^{2}\right\}
&= \mathbf{E}\left\{\left( \sum_{q=1}^{\mathrm{K}}\left(a_{q} \cos{\beta_{q}}- b_{q}\sin{\beta_{q}}\right) \right)^{2}\right\}\\
\end{split}
\end{equation*}
According to multinomial theorem, we can decompose $\left( \sum_{q=1}^{\mathrm{Q}}\left(a_{q} \cos{\beta_{q}}- b_{q}\sin{\beta_{q}}\right) \right)^{2}$ into $\left(a_{q} \cos{\beta_{q}}- b_{q}\sin{\beta_{q}}\right)^{2}$ and $\left(a_{q} \cos{\beta_{q}}- b_{q}\sin{\beta_{q}}\right)\left(a_{p} \cos{\beta_{p}}- b_{p}\sin{\beta_{p}}\right), q \neq p$, repectively.
\begin{equation*}
\begin{split}
&\mathbf{E}\left\{ \left(a_{q} \cos{\beta_{q}}- b_{q}\sin{\beta_{q}}\right)^{2} \right\}\\
% &= \mathbf{E}\left\{ \left(a_{q} \cos{\beta_{q}}\right)^{2}+ \left(b_{q}\sin{\beta_{q}}\right)^{2}-2a_{q} b_{q}\cos{\beta_{q}} \sin{\beta_{q}} \right\}\\
&= \frac{1}{2\pi}\int_{-\pi}^{\pi} \left(\left(a_{q} \cos{\beta}\right)^{2}+ \left(b_{q}\sin{\beta}\right)^{2} -2a_{q} b_{q}\cos{\beta} \sin{\beta}\right)\mathrm{d}\beta\\
&= \frac{1}{2}\left(a_{q}^{2}+b_{q}^{2}\right).
\end{split}
\end{equation*}
% \begin{equation}
% \begin{split}
% &\mathbf{E}\left\{ \left(a_{q} \cos{\beta_{q}}- b_{q}\sin{\beta_{q}}\right)\left(a_{p} \cos{\beta_{p}}- b_{p}\sin{\beta_{p}}\right) \right\}\\
% &= \mathbf{E}\left\{ \left(a_{q} \cos{\beta_{q}}- b_{q}\sin{\beta_{q}}\right)\right\}\mathbf{E}\left\{ \left(a_{p} \cos{\beta_{p}}- b_{p}\sin{\beta_{p}}\right) \right\}\\
% &= 0.
% \end{split}
% \end{equation}
The variance of $\Re\left\{h_{\mathrm{d}}\right\}$ is
\begin{equation}
\begin{split}
    \mathrm{Var}\left\{ \left(\Re\left\{h_{\mathrm{d}}\right\} \right)\right\}
    % &= \mathbf{E}\left\{ \left(\Re\left\{h_{\mathrm{d}}\right\} \right)^{2}\right\}
    % -\left( \mathbf{E}\left\{ \left(\Re\left\{h_{\mathrm{d}}\right\} \right)\right\} \right)^{2}\\
    &= \frac{1}{2}\sum_{q=1}^{\mathrm{N}}\left(a_{q}^{2}+b_{q}^{2}\right) =\frac{1}{2} \sum_{q=\mathrm{1}}^{\mathrm{N}}\left |h_{\mathrm{r},q}   h_{\mathrm{t},q}  \right |^2.
\end{split}
\end{equation}
Similarly,
\begin{equation}
    \mathrm{Var}\left\{ \left(\Im\left\{h_{\mathrm{d}}\right\} \right)\right\}
    = \frac{1}{2} \sum_{q=\mathrm{1}}^{\mathrm{Q}}\left |h_{\mathrm{r},q}   h_{\mathrm{t},q}  \right |^2.
\end{equation}

\subsection{Statistics of the real and imaginary part of the estimated channel}
\label{appB}

\begin{equation}
\begin{split}
        \hat{h} = \sum_{q=1}^{\mathrm{N}}\left |h_{\mathrm{r},q}  \right | \left | h_{\mathrm{t},q}  \right | +\sum_{p=1}^{\mathrm{M}}h_{\mathrm{r},p} h_{\mathrm{t},p} e^{j\beta_{p}} +n
\end{split}
\end{equation}
where $\beta_{p}\sim U\left (-\pi,\pi \right )$, and $n= \sim \mathcal{CN}\left (0,\sigma^{2}_{\mathrm{N}}\right)$. 

$\Re \left\{  \hat{h}  \right\}$ is a Gaussian random variable because it is a sum of constant and independent Gaussian randoms variables. 
\begin{equation}
    \Re \left\{  \hat{h}  \right\} = \sum_{q=1}^{\mathrm{N}}\left |h_{\mathrm{r},q}  \right | \left | h_{\mathrm{t},q}  \right | + \Re \left\{      \sum_{p=1}^{\mathrm{M}}h_{\mathrm{r},p} h_{\mathrm{t},p} e^{j\beta_{p}}       \right\}  +  \Re \left\{n \right\}
\end{equation}
\begin{equation}
    \Im \left\{  \hat{h}  \right\} = \Im \left\{      \sum_{p=1}^{\mathrm{M}}h_{\mathrm{r},p} h_{\mathrm{t},p} e^{j\beta_{p}}       \right\}  +  \Im \left\{n \right\}
\end{equation}
Hence, we have 
\begin{equation}
    \mathbf{E}\left\{\Re \left\{  \hat{h}  \right\} \right\}= \sum_{q=1}^{\mathrm{N}}\left |h_{\mathrm{r},q}  \right | \left | h_{\mathrm{t},q}  \right |,
\end{equation}
 
\begin{equation}
\begin{split}
    \mathbf{Var}\left\{\Re \left\{  \hat{h}  \right\} \right\}
    % &=  \mathbf{Var}\left\{  \Re \left\{      \sum_{p=1}^{\mathrm{M}}h_{\mathrm{r},p} h_{\mathrm{t},p} e^{j\beta_{p}}       \right\} \right\}+ \frac{\sigma^{2}_{\mathrm{N}}}{2}\\
    % &
    = \frac{1}{2} \sum_{q=\mathrm{1}}^{\mathrm{Q}}\left |h_{\mathrm{r},q}   h_{\mathrm{t},q}  \right |^2+ \frac{\sigma^{2}_{\mathrm{N}}}{2}.
\end{split}
\end{equation}
\begin{equation}
    \mathbf{E}\left\{\Im \left\{  \hat{h}  \right\} \right\}= 0,
\end{equation}
\begin{equation}
\begin{split}
    \mathbf{Var}\left\{\Im \left\{  \hat{h}  \right\} \right\}= \frac{1}{2} \sum_{q=\mathrm{1}}^{\mathrm{Q}}\left |h_{\mathrm{r},q}   h_{\mathrm{t},q}  \right |^2+ \frac{\sigma^{2}_{\mathrm{N}}}{2}.
\end{split}
\end{equation}

\subsection{Correlation coefficient between the real and imaginary part of the estimated channel}
\label{appC}

\begin{equation}
\begin{split}
     &\mathbf{E}\left\{\left(\Re\left\{\hat{h}\right\} - \mathbf{E}\left\{ \Re\{\hat{h}\}\right\}\right) \left(\Im\left\{\hat{h}\right\} - \mathbf{E}\left\{ \Im\{\hat{h}\}\right\}\right)\right\} \\
     &= \mathbf{E}\left\{ \Re\left\{\hat{h}\right\}\Im\left\{\hat{h}\right\} \right\} -  \mathbf{E}\left\{ \Re\{\hat{h}\}\right\}  \mathbf{E}\left\{ \Im\{\hat{h}\}\right\}\\
\end{split}
\end{equation}
From Appendices \ref{appA} and \ref{appB}, it is easily to derive that 
\begin{equation}
\begin{split}
     \mathbf{E}\left\{ \Re\left\{\hat{h}\right\}\Im\left\{\hat{h}\right\} \right\} = 0,
\end{split}
\end{equation}
\begin{equation}
\begin{split}
     \mathbf{E}\left\{ \Im\{\hat{h}\}\right\} = 0.
\end{split}
\end{equation}
$\Re\left\{\hat{h}\right\}$ and $\Im\left\{\hat{h}\right\}$ are uncorrelated.

\end{document}